\documentclass[11pt,epsfig]{article}
\usepackage{epsfig}
\newcommand\degr{\,\hbox{$^\circ$}\,}

\newcommand\arcsec{\,\hbox{$^{\prime\prime}$}\,}

\newcommand{\HI}{H{\,\small I}}

\title{\Large {\bf Warps and Rotation Curves in Edge-On Galaxies}}
\author{D. Vergani$^{1}$,  G. Gentile$^1$, R.-J. Dettmar$^2$, G. Aronica$^{2,3}$, U. Klein$^1$\\
\vspace{0.1cm}\\
\normalsize $^1$Radioastronomisches Institut Univ. Bonn, Auf dem H\"ugel 71, D-53121 Bonn, Germany\\
\normalsize $^2$Astronomisches Institut der Ruhr-Universit\"at, Postfach 102148, D-44780 Bochum, Germany\\
\normalsize $^3$Observatoire de Marseille, 2 Place Le Verrier, F-13248 Marseille Cedex 04, France\\
}
\date{}
\begin{document}
\maketitle

\makeatletter
\def\ps@plain{\let\@mkboth\gobbletwo
\def\@oddhead{}\def\@oddfoot{\hfil\tiny
``ALMA extragalactic and Cosmology Science Workshop on Dark Matter'';
Observatoire de Bordeaux, France, 22-24 May 2002}
\def\@evenhead{}\let\@evenfoot\@oddfoot}
\makeatother

\begin{abstract}\noindent

We present our investigation on the effect of warps on the extraction of rotation curves in edge-on galaxies. 
The method to derive the rotation curve from the position-velocity diagram in warped edge-on systems yields 
underestimated velocities, and the tilted-ring model is not reliable in highly inclined, poorly resolved galaxies. 
In a warped system the kinematical major axis is different from the optical major axis. While this is generally a 
limit in optical slit spectroscopy, in the \HI\ emission which extends far from the optical body where self-gravity 
is weaker and the effect of warping is more pronounced, this represents a severe effect to be considered in the 
procedure to extract the rotation curve.

We propose a new approach to extract the rotation curve in highly inclined, warped galaxies. Based on this method 
we are able to trace accurately the frequency of peculiarities in our sample of Thick Boxy Bulge (TBB) galaxies. 
We report an increasing trend of kinematical lopsidedness from spheroidal bulge galaxies towards TBB galaxies.

Concerning the question whether interactions contribute significantly to the bar formation and to the subsequent 
evolution in a box/peanut (b/p) structure, we confirm these theoretical predictions.
Based on our sample, galaxy interaction is the likely formation mechanism to trigger bars in TBB galaxies.

\end{abstract}

\section{Introduction}

A significant fraction of bulges -- up to 50\% -- deviate from the classical spheroidal shape, 
and display excess light above and below the galactic disks, resembling a boxy or peanut shape 
when viewed edge-on.
For this class of bulges there is general agreement that the box/peanut (b/p) shape of the light 
distribution is related to a bar. 
This scenario is strongly supported by observations of the stellar kinematics (Bureau \& Freeman 
1999) as well as by more recent N-body simulations (Athanassoula et al. 2002).
However, the origin of such bars is still not well understood (Sellwood 2000).

At present, two scenarios are widely assumed for the bar formation.
The first one predicts bar formation through the spontaneous instabilities in 
relatively cool and rotationally supported disks, while in the other one the bar is triggered by 
interactions. In fact, interactions both in the form of tidal interaction and direct interaction 
(minor merger) can provide a very efficient mechanism to trigger instabilities in the disks, and 
initiate or speed up bar formation. Once a bar is formed, it is free to evolve in a b/p structure 
due to buckling and thickening (Noguchi 1987; Gerin, Combes \& Athanassoula 1990; Mihos et al. 1995).

The objects of our investigation are b/p bulge galaxies, which however show remarkable differences 
compared to the classical b/p-shaped bulge galaxies in the prominence and thickness of their 
b/p structures. Because of these morphological features, we will refer to them as Thick Boxy Bulges 
galaxies (TBBs, L\"utticke et al. 2000). 

If interactions, spontaneous instabilities, or intermediate mechanisms are the reasons for the origin 
and the evolution of these TBB galaxies they may leave different traces which should be observed. 

For this purpose we have selected a sample of 8 TBB galaxies, and we are investigating the likely 
formation mechanism of this class of objects. The galaxies  were selected from the Third Reference 
Catalogue of Bright Galaxies (de Vaucouleurs et al. 1991) on the basis of their orientation (nearly 
edge-on) and their diameters. The diameters of the galaxies are constrained to be larger than 2$'$ 
at the B$_{25}$ isophote on the Digital Sky Survey (DSS).

Based on the fact that interaction events produce a highly asymmetric distribution and complex
kinematics of the gaseous component, large deviations from symmetry in the morphology and kinematics 
will indicate a galaxy which is strongly interacting with nearby companions.
In order to discern different bar formation scenarios (spontaneous instabilities or bar triggered by 
interactions) one of the diagnostics is the frequency of peculiarities of the TBB galaxies compared 
to spheroidal galaxies.

We are studying the statistics of warps, their shapes as well as lopsidedness.
The latter is investigated both in the kinematics (comparing the approaching and receding side of 
the rotation curve) and in the density distribution (comparing the mass included in the 
receding/approaching side) using 21-cm line VLA and ATCA data.

\section{Rotation curves in our investigation}

Since our investigation has as input the rotation curve from which we will derive the mass and the 
kinematical differences between the two sides in each galaxy, we have to find the most reliable and 
accurate method to obtain the rotation curve. 
Unfortunately, the procedure to derive the rotation curve from highly inclined galaxies is far from 
being simple. It gives rise to severe underestimates of the velocities if conducted with a classical 
approach. This is principally due to the integration along the line of sight of a large portion of 
the disk. Nevertheless, this study has to be performed in edge-on systems since the b/p component is 
a vertical structure observable only in such inclined galaxies. 

Using the tilted-ring model (Begeman 1987) we have the advantage to determine several kinematical 
parameters from initial estimated values through an iterative fit to the velocity field.
However, in our objects the combination of the edge-on orientation (75\degr -- 90\degr) and the small 
angular size often yield a final poor sampling in every ring of the tilted model (less then 15 points). 
In this case, the fit fails or produces results for the rotation curve with poor confidence.
Moreover, the velocity field derived as an intensity-weighted mean velocity (first-moment analysis) or 
by fitting a single-peaked Gaussian to the profiles do not provide a good estimate for the radial velocity, 
which was actually expected since velocity profiles fitted by a symmetric distribution are not reliable.
In edge-on galaxies we are faced with the inconvenience that the beam intercepts a large portion of the 
disk, the effect of the inclination, and the peculiarities of the galaxies (extra-planar emission, 
``beards'', in/outflow) which produce a tail of the distribution at lower velocities, towards the systemic 
velocity.

In order to reduce this effect, the rotation curve could be extracted as outlined in Sancisi \& Allen (1979). 
The velocity is fitted by half a Gaussian to the extreme edge of the velocity profile at each position along 
the major axis. We corrected the velocities for several effects (random motions, instrumental broadening, etc.)
as explained by Gentile et al. (2002).

The advantage of this method is evident. We do not expect to have a single-peaked Gaussian in every velocity 
profile since the profiles are not symmetric, but with this method we utilize the most useful half (the 
external one) of a Gaussian, avoiding the fit over the broadened half owing to the lower velocities.

The disadvantage starts when we have to investigate warped systems (50\% of all spiral galaxies) and the 
position-velocity diagram (PVD) is along a fixed position angle, while in such systems both the line of nodes 
and the kinematical axis are changing with radius.
The effect of warps on the procedure to extract rotation curves over the PVD in edge-on systems is quite severe. 
Moreover, the phenomenon of warping is quite common both in the \HI\ emission (Bosma 1991) and at optical 
wavelengths (Briggs 1990). 

We might limit our analysis to the region which is not affected by this perturbation, but the disadvantage is 
a loss of information. Alternatively, we could make a new approach for the extraction of the rotation curve in 
warped and highly inclined galaxies which takes into account the variation of the line of nodes.

\section{The effect of the warp on the rotation curve}

Figure 1 shows several simple models of a velocity field with a given rotation curve where -- 
applying an increasing variation of the position angle -- we extract the rotation curve in two extreme cases.

The first column represents an unwarped velocity field at the top, then ones with warps of 2\degr, 5\degr, 
7\degr\ and 10\degr, respectively, ranging from 100\arcsec\ to 300\arcsec. In all the models the position angle 
is fixed to 0\degr in the inner 100\arcsec, since usually the warping starts around the edge of the optical disk.
Over these velocity fields we have plotted two slices. The straight green line (slice A) is at the position 
angles of the inner region (optical major axis), while the dashed magenta line (slice B) is at the position 
angle of the last point (2\degr, 5\degr, 7\degr\ and 10\degr). The second column represents the rotation curves 
extracted from slice A, and the third column the rotation curves from slice B. The black line is the real rotation
curve of this system.

Despite the naive model, the effect of warps in deriving the rotation curve over a fixed axis is clear.
We derive two underestimated values: 
\begin{itemize}
\item{ the distance out to which the curve could be traced in the case of slice A (optical major axis);}
\item{ the values of the velocities in both cases (slice A and B).}
\end{itemize}

The velocities are underestimated in the external region, in the case of the rotation over the optical major 
axis (slice A). This might yield an underestimated value of the dynamical mass included out to the last 
observed point. In the case of the slice along the last warped point (slice B), the velocities are underestimated 
in the inner region; this might cause a wrong mass distribution in the central region of a galaxy. 
These effects are more severe for larger warps, as expected.

The method to derive the rotation curve over the PVD in warped edge-on systems yields underestimated rotation 
velocities. The reason is based on the fact that in a warped system the kinematical major axis is neither the 
optical major axis nor the major axis of the last observable point, but rather a {\it flexible} axis.

This is generally a limit in optical slit spectroscopy in which the slit is often oriented a priori along the 
major axis determined from photometry. The kinematical axis might be off by several degrees from the optical 
major axis (due to e.g. dust absorptions). However, this limit is not so severe, due to the fact that often the 
warping starts at the end of the optical body.
In case of \HI, which extends far from the optical body where self-gravity is weaker and the effect of warping 
more pronounced, this represents a severe effect to be considered in the procedure to extract rotation curves.

In order to account for these effects, we propose a new approach to treat the data. We suggest to first find 
the points along the ridge of the warp. The ridge of the warp might be determined fitting a Gaussian distribution 
parallel to the minor axes of the \HI\ total intensity distribution, as outlined by Garc\'{\i}a-Ruiz (2001).  
With this spatial information we derive the rotation curve and the P.A. at each radius from the centre, fitting half a 
Gaussian to the edge of the \HI\ profiles at the positions localized not along a fixed axis, but rather along the 
ridge of the warp (Fig.\,2).

In Fig.\,3 we have plotted an example of one of the observed velocity fields as obtained with this procedure,
called {\it WArped Modified Envelope Tracing} (WAMET) method, and the model velocity field built up with the kinematical 
information (the rotation curve and the P.A. at each radius) derived from this method. The inclination (first guess 
following Verheijen \& Sancisi 1993), centre, and the systemic velocity were obtained by minimizing the residual 
velocity field, which was obtained by subtracting the model from the observed velocity field. We can observe that 
several peculiarities of the velocity field are well reproduced by the model, showing the robustness of the method.

\section{Discussion and Conclusion}

In this report we have explained why the effect of warps cannot be neglected in the procedure to derive \HI\ 
rotation curves, and have stressed why this might affect any further kinematical or dynamical analysis.
We have proposed a different approach to extract the rotation curve in warped, highly inclined galaxies, which can 
now be determined accurately.

Based on this new approach, we are able to accurately trace the frequency of peculiarities in our sample of
TBB galaxies. We report an increasing trend of kinematical lopsidedness, from spheroidal bulge galaxies towards 
TBB galaxies.

Concerning the question whether interactions contribute significantly to bar formation and subsequent b/p structure, 
based on our sample we confirm the theoretical prediction that TBBs have more prominent peculiarities, and interactions 
seem to be the likely formation mechanism of the TBB galaxies. 

Unfortunately, with our limited number of objects we are not able to extend this conclusion to the class of b/p spiral 
galaxies, and this result might be valid only for the local universe. However, this trend exists and further analysis 
might be conducted on a larger sample of objects.

However, for models of galaxy evolution and to investigate the role assumed by interaction events in reshaping galaxy
components (bars, bulges and disks) it is essential that the asymmetries and lopsidedness be well-determined over as 
large a dynamical range and redshifts as possible.
Instruments like the Atacama Large Millimeter Array (ALMA) are needed for good statistics of warping phenomena at 
different redshifts to investigate the still unknown origin of warped galaxies, whether they are shaped under the 
influence of a decoupled dark and luminous matter potential or whether they are dominated by their environment.

\begin{figure}
\vskip -5cm
\centerline{
\epsfig{figure=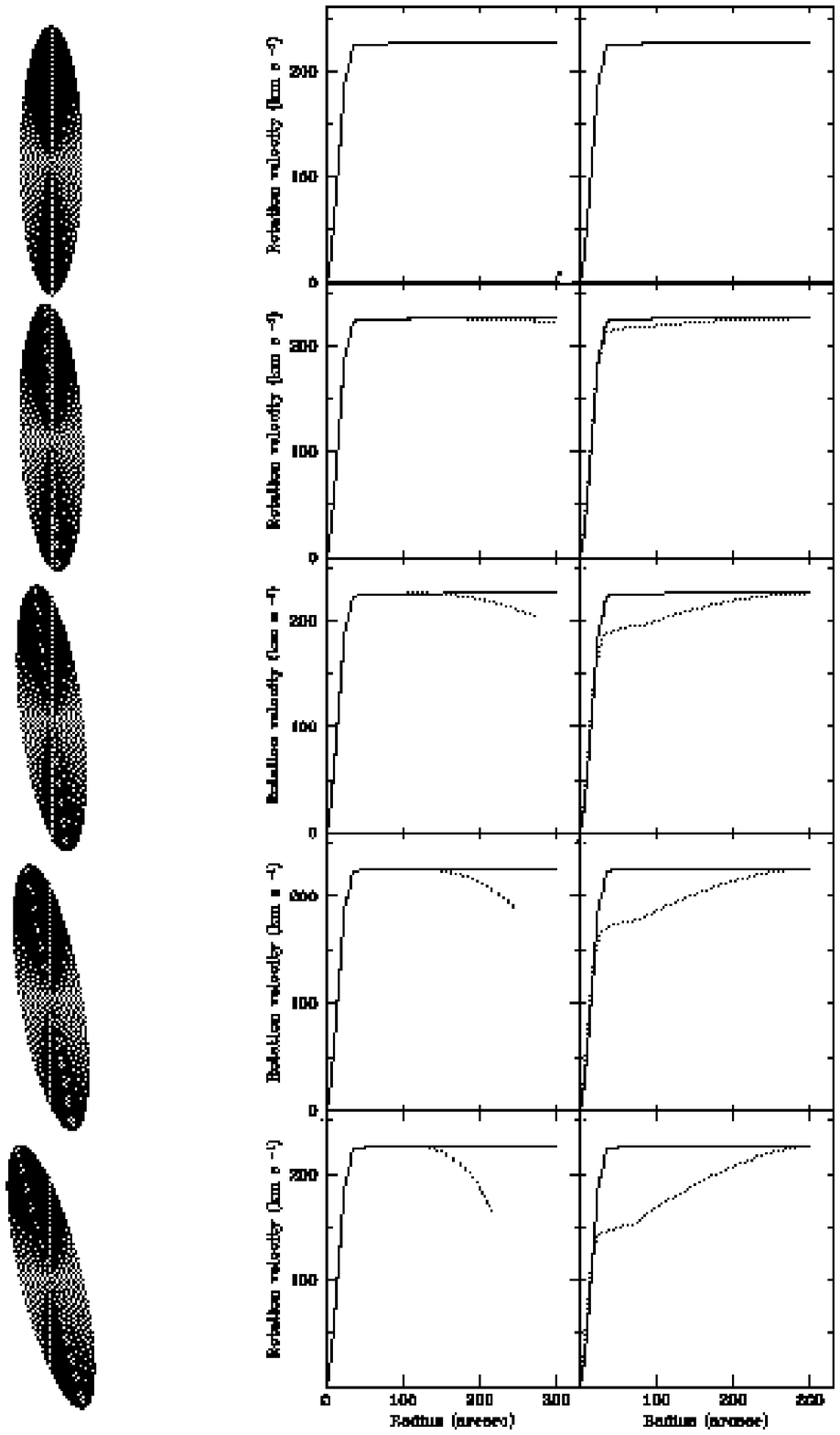,width=20cm,angle=0}
}
\vskip -3cm
\caption{Several simple models of a velocity field with a given rotation curve where -- 
applying an increasing variation of the position angle (see text) -- we extract the rotation curve in two extreme cases
(second and third column).}
\end{figure}

\begin{figure}
\vskip -6cm
\epsfig{figure=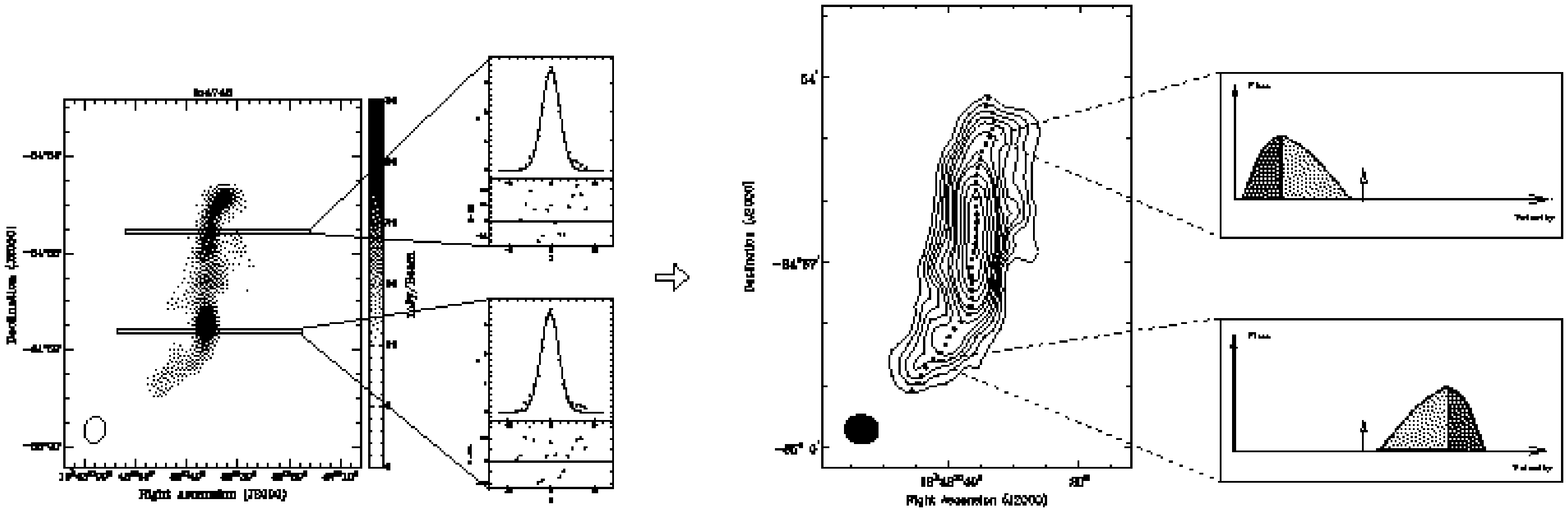,width=13cm,angle=-90}
\vskip -2.5cm
\caption{New approach for the extraction of the rotation curve in warped, highly inclined galaxies called 
{\it WArped Modified Envelope Tracing} (WAMET) method. 
We suggest to find the points along the ridge of the warp, as outlined by Garc\'{\i}a-Ruiz (2001).  
With this spatial information we derive the rotation curve and the P.A. at each radius from the centre, fitting half a 
Gaussian to the edge of the \HI\ profiles at the positions localized along the ridge of the warp.}
\vskip 1cm
\centerline{
\epsfig{figure=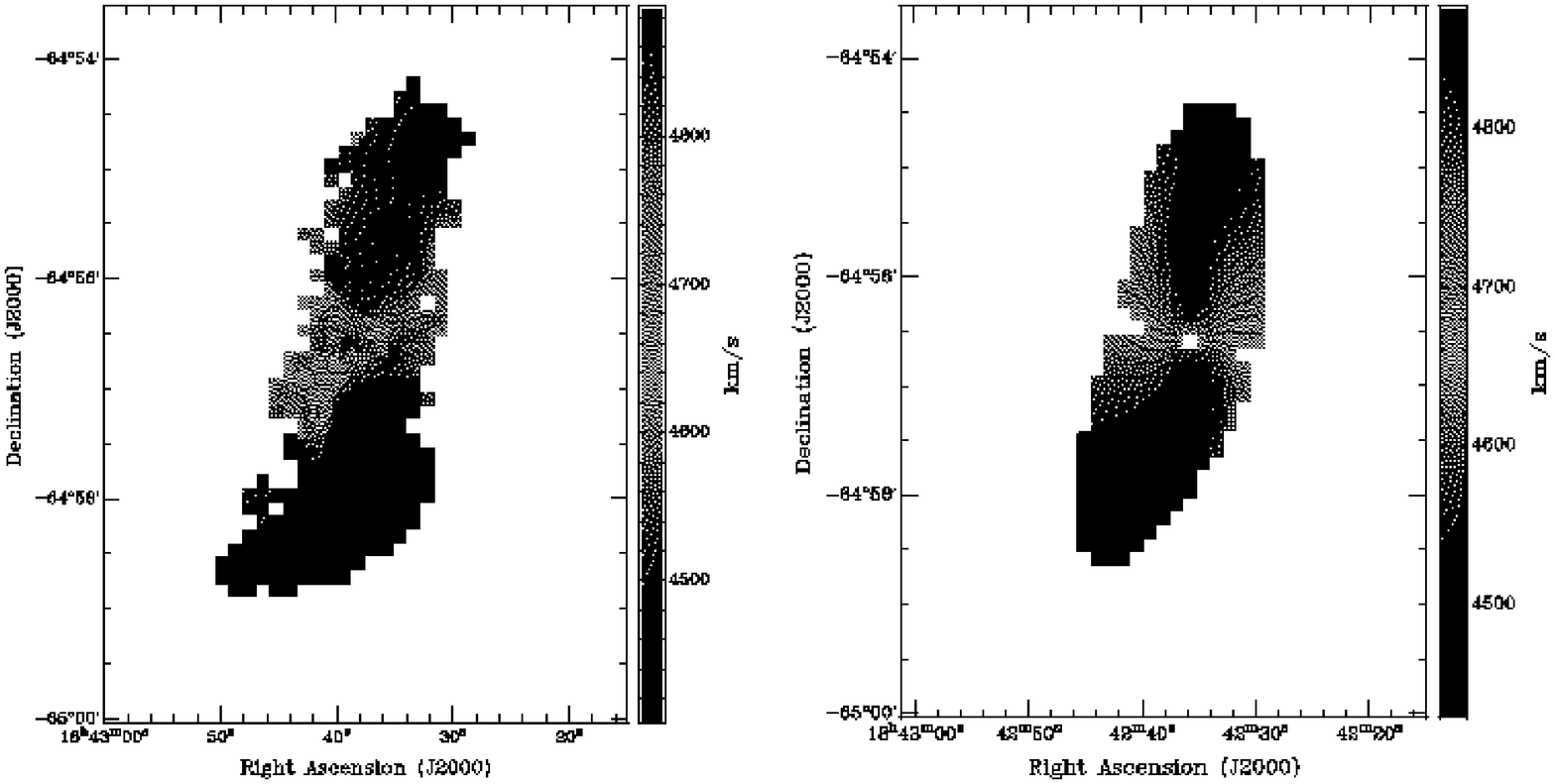,width=11cm,angle=-90}
}
\vskip -1cm  
\caption{An example of one of the observed velocity fields as obtained with the new WAMET procedure is shown 
on the left, together with the model velocity field 
built up with the collected kinematical information (the rotation curve and the P.A. at each radius) on the right. 
The inclination, centre and the systemic velocity were obtained by minimizing the residual velocity field which was 
made by subtracting the model from the observed velocity field. 
We can observe that several peculiarities of the velocity field are well reproduced by the model which show the robustness of
the method.}
\end{figure}

{\vskip 0.4cm
\small {\bf Acknowledgments}}
\vskip 0.1cm
\noindent
We wish to acknowledge Tom Oosterloo and Raffaella Morganti for helpful discussions. 
DV is grateful for financial support of the {\it Deutsche Forschungsgemeinschaft} under number GRK 118 'The 
Magellanic System, Galaxy Interaction and the Evolution of Dwarf Galaxies'.
We thank the Observatoire de Bordeaux for the warm hospitality during the workshop.

\vskip 0.5cm
{\small
\begin{description}{} \itemsep=0pt \parsep=0pt \parskip=0pt \labelsep=0pt
\item {\bf References}

\item Athanassoula E., Misiriotis A., 2002, MNRAS 330, 35
\item Begeman, K.\,G.: 1987, Ph.D. thesis, Univ, Groningen
\item Bosma A., 1991, in S.\,Casertano, P.\,Sackett and F.\,Briggs (eds.), Cambridge University Press, 181
\item Briggs F.\,H., 1990, ApJ 352, 15
\item Bureau M., Freeman K.\,C., 1999, AJ 118, 126
\item de Vaucouleurs G., de Vaucouleurs A., Corwin J.\,R., Buta R.\,J., Paturel G., Fouque P., 1991
\item Garc\'{\i}a-Ruiz, I.: 2001, Ph.D. thesis, Univ. Groningen
\item Gentile G., Vergani D., Salucci P., Kalberla P., Klein U., 2002, in ALMA extragalactic and 
Cosmology Science Workshop on Dark Matter, Bordeaux; http://www.observ.u-bordeaux.fr/public/alma\_workshop/darkmatter 
\item Gerin M., Combes F., Athanassoula L., 1990, A\&A 230, 37
\item L{\"u}tticke R., Dettmar. R.-J., Pohlen M., 2000, A\&AS 145, 405L
\item Mihos J.\,C., 1995, ApJ 428L, 75
\item Nogichi M., 1987, MNRAS 228, 635
\item Sancisi R., Allen R.\,J., 1979, A\&A 74, 73
\item Sellwood, J.\,A., 2000, ASP Conference Series, Vol. 197, p. 3
\item Verheijen, M.\,A.\,W., Sancisi, R.: 2001, A\&A 370, 765
\end{description}
}

\end{document}